\documentstyle[prd,preprint,aps]{revtex}
\begin{document}

\reversemarginpar
\tighten

\title{Algebraic approach to quantum black holes: logarithmic 
corrections to black hole entropy} 
\author{Gilad Gour\thanks{E-mail:~gour@cc.huji.ac.il, 
gilgour@Phys.UAlberta.CA}}
\address{Racah Institute of Physics, Hebrew University of
Jerusalem,\\  Givat Ram, Jerusalem~91904, ISRAEL \\
and\\
Theoretical Physics Institute, Department of Physics, University of Alberta,\\
Edmonton, Canada T6G 2J1}

\maketitle

\begin{abstract} 
The algebraic approach to black hole quantization requires the horizon
area eigenvalues to be equally spaced. As shown previously, for a neutral
non-rotating black hole, such
eigenvalues must be $2^{n}$-fold degenerate if one constructs the black
hole stationary states by means of a pair of creation operators
subject to a specific algebra. We show that the algebra of these two
building blocks exhibits $U(2)\equiv U(1)\times SU(2)$ symmetry, where
the area operator generates the $U(1)$ symmetry. The three generators
of the $SU(2)$ symmetry represent a {\it global} quantum number
(hyperspin) of the black hole, and we show that this hyperspin must be
zero. As a result, the degeneracy of the $n$-th area eigenvalue is
reduced to $2^{n}/n^{3/2}$ for large $n$, and therefore, the
logarithmic correction term $-3/2\log A$ should be added to the 
Bekenstein-Hawking entropy. We also provide a heuristic approach
explaining this result, and an evidence for the existence of {\it two} 
building blocks. 

\end{abstract}

\pacs{PACS numbers:~}

\section{Introduction}

The flurry of activity on the theory of black holes in general
relativity illuminated a very deep and fundamental relationship
between gravitation, thermodynamics and quantum
theory~\cite{Wald99}. The bedrock of this connection is  
the notion of back hole entropy~\cite{bek73}. Bekenstein's
identification of the 
horizon area with black hole entropy has provided a criterion for the
success of any candidate quantum theory of gravity. Today we are
presented with several generic quantum gravity theories 
(superstring and D-brane theory~\cite{string}, loop quantum
gravity~\cite{loop}, canonical 
quantum gravity~\cite{canonical,induced,nearh}) all of which reproduce 
the black hole entropy from a microscopic counting of states. 
Despite this success, there is as yet no generally accepted
theory of quantum gravity; other conditions are needed to
provide further tests on theories which purport to represent quantum
gravity. 

In~\cite{GandM} it is shown that a sufficient
condition for establishing the identification $S_{BH}\propto A$ is
that the complete system, black hole-Hawking radiation, can be
characterized by classical thermodynamics. Furthermore, it is shown
(assuming the first law of black holes mechanics) that any nonlinear 
correction to the black hole entropy must be
quantum mechanical in nature. For example, in a
Reissner-Nordstrom black hole, the dimensionless function 
$\Phi _{T}\equiv(\partial M/\partial Q)_{T}$ (which can be interpreted
as the electric potential on the horizon when the black hole is in 
equilibrium with a surrounding heat bath) can be expanded in powers
of $\hbar$ where the leading term $O(\hbar ^{0})$ is taken to be classical.
It is shown in~\cite{GandM} that higher order terms in $\hbar$
appear if and only if $S''(A)\neq 0$, where $S(A)$ is the entropy
of the black hole and $A$ is the horizon surface area. 
Therefore, the leading order corrections
to the Bekenstein-Hawking entropy (if they exist) will provide a
further insight in the description of quantum effects in black holes.    

Very recently, several authors made an attempt in this direction.   
For example, Kaul and Majumdar~\cite{KandM00} 
(see also~\cite{DKM01,Govinda}) have calculated the lowest 
order corrections to the Bekenstein Hawking entropy in a particular
formulation of quantum loop gravity~\cite{KandM98}. In this
approach, the black hole 
horizon is treated as a boundary spacetime which is quantized within
the ``quantum geometry'' program. They have found a $-3/2\log A$
correction term to the black hole entropy. Carlip~\cite{Carlip00} has
found 
exactly the same correction term for the BTZ black hole
by using the corrected version of the asymptotic Cardy formula. He
has also shown that similar logarithmic corrections appear in all
black holes whose microscopic degrees of freedom are described by an
underlying conformal field theory. Recently, Das, Majumdar and
Bhaduri~\cite{DMB} have computed the leading order corrections to the
entropy of 
any thermodynamic system due to small statistical fluctuations around
equilibrium. These corrections also turn out to be
logarithmic in the horizon area when applied to black holes, and equal
to $-3/2\log A$ when applied to the BTZ black hole. More recently,  
Gupta {\it et al}.~\cite{Gupta} obtain the same correction term by
using a scalar field as a simple 
probe of the background geometry of a massive Schwarzschild black hole.
Here, we shall obtain exactly the same logarithmic correction by using
a simple heuristic scheme, and also by exploiting the algebraic
approach~\cite{Ann_Arbor,BekGour}. We further argue that {\it positive}
logarithmic corrections might also be required in the
Bekenstein-Hawking entropy if the horizon area fluctuations are
included. This may shed some light on why other
authors~\cite{Solo,Fursaev,jy,bss,Car95,MS,Kast97,MakRep98,GourR,OST}
have obtained logarithmic corrections with different coefficients. 

Among the simplest questions which can be asked, and which can also provide 
conditions on a quantum theory of gravity, is what is the nature of the
energy spectrum of a black hole. The remarkable observation that the
horizon area behaves as a classical adiabatic
invariant~\cite{Bek74,BHTrail,Mayo} 
provided the first clue in this direction. In the spirit of Ehrenfest
principle, Bekenstein~\cite{Bek74,Brazil} conjectured that
the horizon area of a non extermal quantum black hole has a discrete
eigenvalue spectrum. He also conjectured that the spacing between area
eigenvalues is uniform~\cite{bek73,Brazil} since the
assimilation of a quantum particle into a Kerr-Newman black hole  
carries a minimal ``cost" $\sim\hbar$ of area increase, which
does not depend on the black hole parameters. The discrete nature of
the eigenvalue spectrum for the horizon area is also supported
by the loop quantum gravity~\cite{loop}, but this last theory suggests
a rather complicated eigenvalue spacing. If the area spectrum is
equally spaced, the classical relation $A=16\pi M^{2}$ ($c=G=1$) for
a Schwarzschild black hole implies the mass spectrum
$M\sim\sqrt{\hbar n}$ for it, where  $n=1,2, \cdots$. This type
of spectrum has subsequently been obtained by many
authors (see the list in~\cite{BekGour} and Ref.~\cite{Krause}). 

The heuristic approach mentioned above is far from giving a complete
description of quantum effects in black holes. In order to illuminate
the inner structure of a black hole, an algebraic approach to
black hole quantization has been developed~\cite{Ann_Arbor,BekGour}. 
Analogously to quantum loop gravity~\cite{loop}, in this
approach one seeks to determine the spectra of relevant observables
in the theory by their algebra. However, in the algebraic approach it
is assumed that each separate black hole state, which one assumes
comes from a discrete set, is created from a ``black hole vacuum'' 
$|vac\rangle$ by a certain ``creation'' operator:
\begin{equation}
|n,s\rangle=\hat{R}_{ns}|vac\rangle.
\end{equation}
Here $|n,s\rangle$ is a one (non-rotating and neutral) black hole
state with area $a_{n}$; $s=0,...,g_{n}-1$ distinguishes between different states
with the same area, where $g_{n}$ is the degeneracy of the said
states. It was shown in~\cite{Ann_Arbor} that the algebra of the 
various $\hat{R}$ operators together with the horizon area observable
(in~\cite{Ann_Arbor} the angular momentum and charge observables are also
included) implies that the spectrum of $\hat{A}$ is equally spaced:
\begin{equation}
 a_{n}=a_{0}n;\qquad n=1,2,3, \cdots,
\label{quant}
\end{equation}  where $a_{0}$ is a positive constant
proportional to $\hbar$.

More recently~\cite{BekGour}, the $\hat{R}$ operators where
constructed from a small number of more fundamental ``building
blocks'' out of which the whole algebra follows. It was assumed for
simplicity that the first area level has two independent quantum
states, say $|1,0\rangle$ and $|1,1\rangle$, so that the corresponding
two building blocks for these states are given by:  
\begin{equation}
\hat{a}\equiv\hat{R}_{11}\quad{\rm
and}\quad\hat{b}\equiv\hat{R}_{12}.
\label{a=R}
\end{equation}  
It was shown that if one builds the $\hat{R}$ operators as products
involving only these two non-commuting operators, the degeneracy of
the area levels is given by the exponential law $g_{n}=2^{n}$. This was
the first time that a formal proof has been given for this law.

The
exponential law $g_{n}=k^{n}$ ($k=2,3,...$) was first proposed by
Mukhanov~\cite{mukh86} (see also Ref.~\cite{bekmukh}) where it was
assumed that the entropy of the black hole corresponds to the
degeneracy of the area levels Eq.~(\ref{quant}). Since then quite a
few heuristic ways of understanding the exponential growth of
degeneracy have been
proposed~\cite{Brazil,bekmukh,DS,wheeler,sorkin,kastrup}. 
One of the simplest views to get the
quantization law Eq.~(\ref{quant}) suggests that the horizon
may be regarded as parceled into $n$ patches of area
$a_0$.  If each can exists in $k$ different quantum states,
then the law $g=k^n$ is immediate~\cite{Brazil}. However, as we shall see
here, if one includes the {\it hyperspin} or the {\it global} quantum
number of these patches, the exponential law (for $k=2$) is
replaced by 
\begin{equation}
g_{n}={n\choose n/2}-{n\choose n/2+1}\approx\frac{2^{n}}{n^{3/2}}
\label{ggg}
\end{equation}     
for $n\gg 1$. We shall obtain this result also in a more formal way
using the algebraic approach where the hyperspin appears as the
generator of the symmetry between the two building blocks in
Eq.~(\ref{a=R}). Furthermore, in the Appendix, we start with $k=3$
(equivalent to three building blocks) and find in this case that 
$g_{n}\approx\frac{3^{n}}{n^{2}}$ which corresponds to a $-2\log n$
correction term.
We shall mention here, that by using a holographic
point of view, Das, Kaul and Majumdar~\cite{DKM01} have obtained the
same result. However, in their formalism the left hand side of
Eq.~(\ref{ggg}) is replaced by the number of $SU(2)$ singlet states
contributing to the entropy of an isolated horizon (which is described
by the boundary degrees of freedom of a three dimensional Chern Simons
theory), and $n$ is replaced by the number of vertices of the network. 

The physical interpretation of the building blocks $\hat{a}$ and
$\hat{b}$ may be elucidated by comparison with other approaches.
It was shown~\cite{DRY} that the algebraic approach and the approach 
involving quantization on a reduced phase space~\cite{BDK01} are
similar. However, it is still not clear how the degeneracy appears
in the later approach. Here, we shall argue that one of
the building blocks is associated with the boundary at infinity 
through the ADM mass $M$ and its conjugate $P_{M}$ 
(the time separation at special infinity), and the other with the
boundary at the horizon via $A$ and its conjugate $\Theta$ (the
``opening angle'', see Ref.~(\cite{CarTet})). This also gives an
argument for setting $k=2$.

The paper is organized as follows: In Sec.~\ref{sec:heuristic}, we
develop a heuristic approach to calculate the $-3/2$ logarithmic
correction term. In Sec.~\ref{sec:interpretation}, we provide a
physical interpretation to the building blocks, by a comparison with
the reduced phase space approach. In Sec.~\ref{sec:algebra}, the
hyperspin is defined within the algebraic approach, and the degeneracy
of the area levels in the hyperspin representation is obtained. We
conclude with a discussion (Sec.~\ref{sec:discussion}). The
Appendix generalizes the results to the case of three building
blocks. 

\section{Heuristic scheme and logarithmic corrections}
\label{sec:heuristic}

Since any correction to the
black hole entropy formula, $S_{\rm bh}=\eta A $, must be quantum
mechanical in nature~\cite{GandM}, the leading order corrections to the black hole
entropy (if they exist) will provide a further insight into the interplay
between general relativity and quantum mechanics. Furthermore, as one reduces
the size of the black hole, these corrections may play an important
role and even change completely the semi-classical description. 

In previous work~\cite{GourR}, we have found a {\it 
positive} logarithmic correction $1/2\log A$ to the Bekenstein-Hawking
entropy by considering the Schwarzschild black hole as a grand
canonical ensemble, with the Hamiltonian (the ADM mass) and the
horizon surface area, separately, as observable parameters. Also
Kastrup~\cite{Kast97}, M\"{a}kel\"{a} and Repo~\cite{MakRep98} and
Obregon, Sabido and Tkach~\cite{OST} obtained exactly the same 
{\it positive} correction. More recently, Major and Setter~\cite{MS}
obtained the correction term $\log A$ using a model which is motivated
by the formulation of geometry in loop quantum gravity (they
also describe the black hole in terms of a grand canonical ensemble).     
All these approaches predict positive corrections due to the area
fluctuations. The area fluctuations {\it increase} the uncertainty, and 
therefore, the number of microstates which describe the black hole. 
However, the logarithmic corrections which
correspond to corrections in the degeneracy $g(n)=k^{n}$ are
not positive. We shall give now a simple heuristic argument, based on the
equally spaced area spectrum, which yields the correction term
$-3/2\log A$ (in section 4 we shall derive this result from the
algebraic approach). This term was obtained by several
authors~\cite{KandM98,DKM01,Govinda,Carlip00,DMB,Gupta} using
a variety of techniques. 

In several models of black holes, it is assumed that the black hole 
consists of elementary components contributing additively to its
area. In quantum loop gravity, these components are described by the
Wilson lines of the Ashtekar's connection $A^{a}_{\mu}$ (see for
example Ref.~\cite{Gambi}). In $M$-theory~\cite{Banks}, the components are
described by the $D_{0}$-branes. Here, we 
consider the equally spaced area spectrum, where the horizon
may be regarded as parceled into $n$ patches (components) of area
$a_{0}$. In this heuristic view, these patches are localized. Thus,
it is assumed that each cell has the same number of quantum states due
to rotational symmetry. For simplicity, we assume that each cell can
exist in 2 different quantum states. 

Since the elementary components (the patches) can exist in two
states, in quantum mechanics they can be described by $1/2$-spin
entities. These are here called ``hyperspins'' to distinguish them from spin;
i.e. the $1/2$-hyperspin which is associated with each patch describes
its two different quantum states. The $1/2$-hyperspin observable associated
with the $i$'th patch is denoted by $\vec{\bf h}^{(i)}\equiv
(\hat{h}_{1}^{(i)},\hat{h}_{2}^{(i)},\hat{h}_{3}^{(i)})$ where the
total hyperspin of a black hole with area $A=a_{0}n$ is given by
\begin{equation}
\vec{{\bf {\cal J}}}=\sum_{i=1}^{n}\vec{\bf h}^{(i)}.
\end{equation}
Since $\vec{{\bf {\cal J}}}$ represents a {\it macroscopic} feature of
the black hole, it should have played an important role also in the
classical description. However, according to Wheeler's no-hair
principle, the Schwarzschild black hole is parametrized only by its
mass. This leaves us no choice but to set the total hyperspin of the
Schwarzschild black hole to zero. 
Furthermore, by analogy with a system of 1/2-spin
particles, it seems reasonable that the energy (mass) should depend on
${\cal J}$. If so, the relation $A=16\pi M^{2}$ holds true only in the
classical limit where ${\cal J}=0$. This motivates the idea that the 
Schwarzschild black hole should be considered as a grand
canonical ensemble~\cite{MS,GourR}.

The number of microstates with area $A=a_{0}n$ is $g(n)=2^{n}$. This
follows from the two quantum states each cell can exist in. However, the  
number of microstates $g(n,{\cal J}=0)$ with area $A=a_{0}n$ and zero
hyperspin ${\cal J}=0$ is smaller. A simple calculation (see
section~\ref{sec:algebra}) shows that 
\begin{equation}
g(n,{\cal J}=0)=\frac{1}{1+1/2n}{n \choose 1/2n}.
\label{dege}
\end{equation}
Note that $n$ must be even since ${\cal J}=0$.
Using Stirling's formula, we find in the classical limit
$n\rightarrow\infty$ that 
\begin{equation}
g(n,{\cal J}=0)\approx 2^{n}/n^{3/2}.
\end{equation}
This yields a correction term $-3/2\log n$ to the Bekenstein-Hawking
entropy. Moreover, Eq.~(\ref{dege}) shows clearly that as one reduces
the size of the black hole, the semi-classical description is
completely changed. For example, an elementary black hole with area
$A=2a_{0}$ has zero entropy.

In conclusion, logarithmic corrections to the black hole entropy may
be classified into two types: {\it positive} corrections, which appear
due to the area fluctuations, and {\it negative} corrections, which
appear as a result of corrections to the number of microstates that describe
a black hole with a {\it definite} horizon area.    

\section{A physical interpretation of the building blocks }
\label{sec:interpretation}

In previous work~\cite{BekGour}, it was suggested that the algebra of
black holes observables, can be constructed from two basic operators 
({\it building blocks}) $\hat{a}$ and $\hat{b}$, satisfying the
following commutation relation:
\begin{eqnarray}
\;[\hat{A},\hat{a}] & = & a_{0}\hat{a}
\;,\;\;\quad[\hat{A},\hat{b}]=a_{0}\hat{b}\label{algebra1}\\
\;[\hat{a}^{\dag},\hat{a}] & = & [\hat{b}^{\dag},\hat{b}]=1+\alpha\hat{A}
\equiv 1+w\hat{N}\label{algebra2}\\
\; [\hat{a}^{\dag},\hat{b}] & = &
[\hat{b}^{\dag},\hat{a}]=0 
\label{algebra}
\end{eqnarray}
where $\hat{N}\equiv\hat{A}/a_{0}$ is a dimenssionless number operator.
The commutation relation between $\hat{a}$ and $\hat{b}$ may be complicated
but it does not matter for our analysis because the degeneracy is determined
by relations (\ref{algebra1}-\ref{algebra}) only.
It was shown that this algebra leads to exactly $2^{n}$ independent states
with the same area eigenvalue $a_{0}n$. This degeneracy corresponds to
the entropy of the black hole; in the next section the logarithmic
correction term $-3/2\ln n$ will be obtained. 

The physical interpretation of the building blocks $\hat{a}$ and
$\hat{b}$ may be found by comparison with other approaches. For
example, in~\cite{DRY} it was shown that the algebraic
approach and the approach involving quantization on a reduced phase
space of collective coordinates of a black hole are similar. 
In the reduced phase space approach~\cite{BDK01}, one starts, still
classically, with an effective action of the form~\cite{LMK}   
\begin{equation}
I=\int dt\left(P_{M}\dot{M}-H(M)\right)  
\label{action}
\end{equation}
where $M$ is the mass and $P_{M}$ its conjugate momentum (here we
consider non-rotating, neutral black holes). The
conjugate momentum $P_{M}$ has the interpretation of the difference
between the Schwarzschild times at left and right
infinities~\cite{Kuchar94} and,  
according to Euclidean quantum gravity~\cite{Hawking}, it is
periodic with a period which is inverse the Hawking temperature
$T_{H}$. The canonical transformation 
\begin{eqnarray}
X & = & 2M\cos(2\pi P_{M}T_{H}/\hbar)\nonumber\\
\Pi _{X} & = & 2M\sin(2\pi P_{M}T_{H}/\hbar) 
\label{canon}
\end{eqnarray}
implies that the horizon area $A=16\pi M^{2}$ can be written in the
form $A=4\pi (X^{2}+\Pi _{X}^{2})$. Thus, after quantization
($X\rightarrow\hat{X},\;\Pi _{X}\rightarrow\hat{\Pi}_{X}$ where
$[\hat{X},\hat{\Pi}_{X}]=i\hbar$) the area observable exhibits an
equally spaced area spectrum. 
Another way to obtain this result, is to perform the canonical
transformation 
\begin{equation}
(M, P_{M})\;\;\rightarrow\;\;(A/8\pi,\Theta)
\label{th}
\end{equation}
where $\Theta\equiv\kappa P_{M}$ and $\kappa$ is the surface gravity. Note
that $\Theta$ has periodicity of $2\pi$ and that it is the canonical
conjugate of $A/8\pi$. Thus, after quantization $A/8\pi$ is
represented by a number operator.  

One can identify one of the building
blocks, say $\hat{a}$, with the creation operator
$1/\sqrt{2\hbar}(\hat{X}-i\hat{\Pi}_{X})$. However, how is the other
building block included? In other words, where is the black hole
entropy in all this? It was suggested in~\cite{DRY} to
include another operator representing the internal degrees of
freedom (similar to the ``secret'' operator described in~\cite{Gour2}).
In the following, we present another way to include the second building
block.
  
Carlip and Teitelboim~\cite{CarTet} have shown that
the standard (Euclidean) action principle for the gravitational field
implies that for spacetimes with black hole topology, the opening
angle $\Theta$ at the horizon and the horizon area $A/8\pi$ are canonical 
conjugates. They have argued that the boundary term at the horizon, 
$-A\delta\Theta$, should supplement the canonical action just as
the boundary term at infinity, $M\delta T$, does ($T\equiv P_{M}$,
the time separation at spatial infinity, is the canonical conjugate to
the ADM mass). Classically, $A=16\pi M^{2}$ and $\Theta=\kappa P_{M}$
(see the last paragraph of this section);
however, the above argument motivates us to consider (initially) $M,\;
P_{M},\;A/8\pi$ and  $\Theta$ as if they were four independent
parameters. Only at the end, after quantization, do we impose the
classical relation. This technique has appeared also in our grand
canonical approach~\cite{GourR}. Adding to Eq.~(\ref{canon}) the canonical
transformation $Y=\sqrt{A/4\pi}\cos\Theta$ and $\Pi_{Y}=\sqrt{A/4\pi}\sin\Theta$,
implies that
\begin{eqnarray}
16\pi M^{2} & = & 4\pi (X^{2}+\Pi _{X}^{2})\nonumber\\
A & = & 4\pi (Y^{2}+\Pi _{Y}^{2}).
\end{eqnarray}
Thus, after quantization, the two ``building blocks'' $\hat{a}\equiv
1/\sqrt{2\hbar}(\hat{X}-i\hat{\Pi}_{X})$ and $\hat{b}\equiv
1/\sqrt{2\hbar}(\hat{Y}-i\hat{\Pi}_{Y})$ raise $16\pi
\hat{M}^{2}$ and $\hat{A}$, respectively, by one unit. Since $\hat{A}$ and $\hat{M}$ are
considered independent, so are $\hat{X}$ and $\hat{Y}$, and thus
$[\hat{a},\hat{b}]=0$. However, since classically $A$ and $M$ are not
independent, one must impose a certain condition relating $\hat{A}$
and $\hat{M}$ (this is
analogous to the requirement of zero hyperspin). After imposing this
condition $\hat{a}$ and $\hat{b}$ no longer commute, in harmony with
the algebraic approach~\cite{BekGour}. In this way we have obtained
two building blocks: one is associated with the boundary at infinity
through $M$ and $P_{M}$, and the other with the boundary at the
horizon via $A$ and $\Theta$. Thus, by a comparison with the reduced
phase space approach, we have found that there are exactly two building
blocks, i.e. $k=2$.

It is interesting to note that the ``opening angle'' defined
classically by Carlip and Teitelboim is related to $M$ and $P_{M}$ by
$\Theta=\kappa P_{M}$~\footnote{In Eq.(~\ref{th}) $\;\kappa P_{M}$ is
  denoted by $\Theta$; here we show that it is actually the ``opening
  angle'' defined by Teitelboim and collaborators}. 
This follows from   
\begin{equation}
1=\frac{1}{8\pi}\{A,\Theta\}_{M,P_{M}}\equiv\frac{1}{8\pi}\left(\frac{\partial A}{\partial M}
\frac{\partial \Theta}{\partial P_{M}}-\frac{\partial A}{\partial P_{M}}
\frac{\partial \Theta}{\partial M}\right),
\label{comat}
\end{equation}
where the Poisson brackets are taken with respect to the canonical
coordinates $(M,P_{M})$. Now, since $A=16\pi M^{2}$ is independent of
$P_{M}$, Eq.~(\ref{comat}) implies that 
$\partial \Theta /\partial P_{M}=8\pi\partial M /\partial A =\kappa$. 
Hence, $\Theta=\kappa P_{M}$, and since $\Theta$ has a period $2\pi$,
the Euclidean time $P_{M}$ has a period $2\pi/\kappa$. In this way
we derive the Hawking temperature, which according to finite
temperature quantum field theory is equal to $\hbar$ divided by the
Euclidean period $2\pi/\kappa$. 

\section{The hyperspin within the algebraic approach}
\label{sec:algebra}

In previous work~\cite{BekGour}, it was shown that the algebra of
the two building blocks (see Eq.~(\ref{algebra1}-\ref{algebra})) 
leads to $2^n$-fold
degeneracy of the $n$-th area level.
Thus, states with the same area eigenvalue $na_{0}$  
could be represented in a binary form:
\begin{equation} |x_{1}x_{2}\cdots
x_{n}\rangle\!\rangle\equiv\hat{x}_{1}\hat{x}_{2}\cdots
\hat{x}_{n} |{\rm vac}\rangle
\label{states1}
\end{equation} where $x_{i}=0\;{\rm or}\;1$  and correspondingly
$\hat{x}_{i}$ is either
$\hat{a}$ or $\hat{b}$ ($i=1,2,..,n$). As we shall see in the
following, not all the states in Eq.~(\ref{states1}) correspond to zero
hyperspin. 

The commutation relations in Eq.~(\ref{algebra1}-\ref{algebra}) are invariant
under the transformation:
\begin{equation}
\pmatrix{\hat{a} \cr
\hat{b} \cr}
\;\;\rightarrow\;\;
\pmatrix{u_{11} & u_{12} \cr
u_{21} & u_{22} \cr}
\pmatrix{\hat{a} \cr
\hat{b} \cr}\;\equiv\;
{\bf u}\pmatrix{\hat{a} \cr
\hat{b} \cr}\;,
\end{equation}
where ${\bf u}\in U(2)$ is a 2-dimensional unitary matrix (here boldface
indicates matrices).
The group $U(2)$ can be written as a direct product of
$U(1)$ and $SU(2)$, i.e. $U(2)=U(1)\times SU(2)$. 
It follows from Eq.~(\ref{algebra1}) that
\begin{eqnarray}
\exp(-i\theta\hat{N})\hat{a}\exp(i\theta\hat{N}) & = & 
\exp(-i\theta)\hat{a}\nonumber\\
\exp(-i\theta\hat{N})\hat{b}\exp(i\theta\hat{N}) & = & 
\exp(-i\theta)\hat{b}.
\end{eqnarray}
Hence, the generator
of the $U(1)$ symmetry is the area operator $\hat{N}$ itself.
We shall denote the generators of the $SU(2)$ symmetry by
$\vec{\cal J}=(\hat{\cal J}_{1},\hat{\cal J}_{2},\hat{\cal J}_{3})$ and we shall 
call $\vec{\cal J}$ the (total) $hyperspin$ of the black hole. 
Thus, the hyperspin arise as a result of the symmetry between the two
building blocks $\hat{a}$ and $\hat{b}$. Since the hyperspin generates
the $SU(2)$ symmetry,
\begin{equation}
[\hat{\cal J}_{i},\hat{\cal J}_{j}]=i\varepsilon_{ijk}\hat{\cal J}_{k}.
\label{hyp}
\end{equation}

The commutation 
relations between $\hat{\cal J}_{k}$ $(k=1,2,3)$ and $\hat{a}$ (or $\hat{b}$), 
can be deduced from the definition of $\vec{\cal J}$ as the generators
of the symmetry group. That is,
\begin{equation}
\pmatrix{
\exp(-i\theta \hat{\cal J}_{k})\hat{a}\exp(i\theta \hat{\cal J}_{k}) \cr
\exp(-i\theta \hat{\cal J}_{k})\hat{b}\exp(i\theta \hat{\cal J}_{k}) \cr}
=\exp(i\theta {\bf \sigma}_{k})
\pmatrix{\hat{a} \cr \hat{b} \cr}\;,
\label{hyper}
\end{equation}
where $\{\sigma_{k}\}$ are the three Pauli matrices.
Now, taking the derivative with respect to $\theta$ in both sides of 
Eq.~(\ref{hyper}), and then setting $\theta =0$ we find that:
\begin{eqnarray}
&{}& [\hat{\cal J}_{1},\hat{a}]=\frac{1}{2}\hat{b}\;\;\;,\;\;\;
[\hat{\cal J}_{1},\hat{b}]=\frac{1}{2}\hat{a}\nonumber\\
&{}& [\hat{\cal J}_{2},\hat{a}]=\frac{i}{2}\hat{b}\;\;\;,\;\;\;
[\hat{\cal J}_{2},\hat{b}]=-\frac{i}{2}\hat{a}\nonumber\\ 
&{}& [\hat{\cal J}_{3},\hat{a}]=\frac{1}{2}\hat{a}\;\;\;,\;\;\;
[\hat{\cal J}_{3},\hat{b}]=-\frac{1}{2}\hat{b}.
\label{commu}
\end{eqnarray}
These commutation relations determine the hyperspin uniquely. 
For example, the third component of the hyper-spin
measure the difference between the number of $a$'s and the number
of $b$'s divided by two; i.e.
\begin{equation}
\hat{\cal J}_{3}=\frac{1}{2}(\hat{N}_{a}-\hat{N}_{b})\;,
\label{j3}
\end{equation}
where $\hat{N}_{a}\equiv\hat{N}-\hat{N}_{b}$ and $\hat{N}_{b}$ is
defined by 
\begin{equation} 
\hat{N}_{b}|x_{1}x_{2}\cdots x_{n}\rangle\!\rangle 
=\left(\sum_{i=1}^{n}x_{i}\right)
|x_{1}x_{2}\cdots x_{n}\rangle\!\rangle\;.
\end{equation}

\subsection{The hyperspin representation}

The $2^{n}$ states defined in Eq.~(\ref{states1}) span the Hilbert
space of all states representing a black hole with an area
$a_{0}n$. Here we would like to classify these states according to
their hyperspin. Henceforth 
the states in the hyperspin representation will be denoted by
$|n,{\cal J},{\cal J}_{3},s\rangle$. 
Eq.~(\ref{hyp}) implies that ${\cal J}$ and ${\cal J}_{3}$
must be integers or half integers. Further, according to
Eq.~(\ref{j3}) the maximum possible value of ${\cal J}_{3}$ (and
therefore of ${\cal J}$) is $n/2$. Thus, ${\cal J}=0,1,2,...,n/2$ (if
$n$ is even) or ${\cal J}=1/2,3/2,...,n/2$ (if $n$ is odd) and 
${\cal J}_{3}=-{\cal J},-{\cal J}+1,...,{\cal J}$. 
The quantum number 
$s=0,1,...,g(n,{\cal J},{\cal J}_{3})-1$ distinguishes between states
with the same area and the same hyperspin;
$g(n,{\cal J},{\cal J}_{3})$ is the number of independent states with
the same area level and the same hyperspin (later we shall
see that $g(n,{\cal J},{\cal J}_{3})$ does not actually depend on ${\cal J}_{3}$).
Thus, the set of states $\{|n,{\cal J},{\cal J}_{3},s\rangle\}$
represents an orthonormal basis for the black hole Hilbert space. 

We shall mention here that Eq.~(\ref{j3}) also
implies that for a given ${\cal J}$ the minimum $n$ is 
$2{\cal J}$. Therefore, the area spectrum can be written as
\begin{equation}
A_{l,{\cal J}}=a_{0}(l+2{\cal J})\;,
\label{areas}
\end{equation}
where $l=0,1,2,...$. However, since the hyperspin of a black hole
which was formed by a gravitational collapse is zero (see
Sec.~\ref{sec:heuristic} and Sec.~\ref{sec:discussion}) this spectrum
coincide with Eq.~(\ref{quant}).  

The $n=0$ area level (the vacuum state $|vac\rangle$) has zero
hyperspin and it is denoted here by $|0,0,0,0\rangle$.
The $n=1$ area level has two independent states:
$|1,1/2,1/2,1\rangle\equiv|0\rangle\!\rangle\equiv\hat{a}|vac\rangle$
and
$|1,1/2,-1/2,1\rangle\equiv|1\rangle\!\rangle\equiv\hat{b}|vac\rangle$. 

The $n=2$ area level has 4 independent (orthonormal) states. 
One state with zero hyperspin and three with ${\cal J}=1$:
\begin{eqnarray}
|2,1,-1,1\rangle & = & \frac{1}{\sqrt{2+w}}|11\rangle\!\rangle\nonumber\\
|2,1,0,1\rangle & = & \frac{1}{\sqrt{4+2w}}
\left(|01\rangle\!\rangle+|10\rangle\!\rangle\right)\nonumber\\
|2,1,1,1\rangle & = & \frac{1}{\sqrt{2+w}}|00\rangle\!\rangle\nonumber\\
|2,0,0,1\rangle & = & \frac{1}{\sqrt{2w}}
\left(|01\rangle\!\rangle-|10\rangle\!\rangle\right).
\label{states}
\end{eqnarray}
Note that $s=1$ in all the states in Eq.~(\ref{states}) since
$g(n=2,{\cal J},{\cal J}_{3})=1$ for all ${\cal J}=0,1$ ${\cal
J}_{3}=1,0,-1$. However, this is no longer true 
for $n>2$. For example, using the commutation relations~(\ref{commu}),
it can be shown that the two independent states
$|010\rangle\!\rangle-|001\rangle\!\rangle$ and 
$|100\rangle\!\rangle-|010\rangle\!\rangle$ 
both have $n=3,\;{\cal J}=1/2$ and ${\cal J}_{3}=1/2$.
We shall turn now to calculate explicitly the number of independent
states with the same $n$ and ${\cal J}$.

\subsection{Degeneracy in the hyperspin representation}

First, we shall show that $g(n,{\cal J},{\cal J}_{3})$ is {\it not} a function
of ${\cal J}_{3}$. In order to prove that, consider the $g(n,{\cal
J},{\cal J}_{3})$ independent states 
$$\{|n,{\cal J},{\cal J}_{3},s\rangle\}_{s=0,1,...,g(n,{\cal J},{\cal
J}_{3})-1}$$.  
As mentioned above, 
states with different $s$ are orthogonal. If ${\cal J}_{3}\neq\pm {\cal J}$
then $\hat{\cal J}_{\pm}|n,{\cal J},{\cal J}_{3},s\rangle\neq 0$ for
all $s=0,1,...,g(n,{\cal J},{\cal J}_{3})-1$ , where 
$\hat{\cal J}_{\pm}\equiv \hat{\cal J}_{1}\pm i\hat{\cal J}_{2}$. 
Now, the third component of the hyperspin of all the $g(n,{\cal
J},{\cal J}_{3})$ states  
$$\{\hat{\cal J}_{+}|n,{\cal J},{\cal J}_{3},s\rangle\}_ {s=0,1,...,
g(n,{\cal J},{\cal J}_{3})-1}$$
is ${\cal J}_{3}+1$. Furthermore, these states are also orthogonal because 
\begin{eqnarray}
\langle n,{\cal J},{\cal J}_{3},s'|\hat{\cal J}_{+}^{\dag}\hat{\cal J}_{+}|n,{\cal J},{\cal J}_{3},s\rangle
& = & \langle n,{\cal J},{\cal J}_{3},s'|(\hat{\cal J}^{2}-\hat{\cal J}_{3}^{2}-\hat{\cal J}_{3})
|n,{\cal J},{\cal J}_{3},s\rangle\nonumber\\
& = & ({\cal J}({\cal J}+1)-{\cal J}_{3}^{2}-{\cal J}_{3})
\langle n,{\cal J},{\cal J}_{3},s'|n,{\cal J},{\cal J}_{3},s\rangle=0
\end{eqnarray}
for $s\neq s'$. In the same way it can be shown that the $g(n,{\cal J},{\cal J}_{3})$
states $\{\hat{\cal J}_{-}|n,{\cal J},{\cal J}_{3},s\rangle\}$ are all orthogonal with
${\cal J}_{3}-1$. Hence, $g(n,{\cal J},{\cal J}_{3})
=g(n,{\cal J},{\cal J}_{3}+1)=g(n,{\cal J},{\cal J}_{3}-1)\equiv g(n,{\cal J})$.
Note that $g(n,{\cal J})$ counts the number of states with the same $n$, ${\cal J}$
{\it and} ${\cal J}_{3}$, even though it does not depend on ${\cal J}_{3}$.

According to Eq.~(\ref{j3}) 
\begin{equation}
N_{b}=\frac{1}{2}n-{\cal J}_{3}
\end{equation}
because $N_{a}+N_{b}=n$. Thus, 
the number of independent states with the same area $n$ and the same
${\cal J}_{3}$ is ${n\choose N_{b}}={n\choose n/2-{\cal J}_{3}}$.
For example, the maximum value of ${\cal J}_{3}$ is
$n/2$, and in this case $N_{a}=n$ and $N_{b}=0$. Thus, there is only
one independent state with area $n$ and with ${\cal J}_{3}=n/2$; that
is the state $|00...0\rangle\!\rangle$. This implies that   
$g(n,{\cal J}=n/2)={n\choose 0}=1$ because $g(n,{\cal J}=n/2)\equiv 
g(n,{\cal J}=n/2,{\cal J}_{3})$ is independent of ${\cal J}_{3}$ and 
for ${\cal J}=n/2$ we can set (for convenience) ${\cal J}_{3}=n/2$. 

The number of states with ${\cal J}_{3}=n/2-1$ ($N_{b}=1$) is equal to
${n\choose 1}$. Now, $g(n,{\cal J}=n/2)$ is the number of states with 
${\cal J}=n/2$ and a particular ${\cal J}_{3}$; in particular it 
is also the number states with
${\cal J}=n/2$ and ${\cal J}_{3}=n/2-1$ (remember that 
$g(n,{\cal J}=n/2)\equiv g(n,{\cal J}=n/2,{\cal J}_{3})$ for any
${\cal J}_{3}$).
Therefore,
\begin{equation} 
g(n,{\cal J}=n/2-1)={n\choose 1}-g(n,{\cal J}=n/2)
={n\choose 1}-{n\choose 0}.
\end{equation}

Similarly, in order to obtain $g(n,{\cal J}=n/2-2)$, we have to
subtract from the ${n\choose 2}$ states with the same 
${\cal J}_{3}=n/2-2$ the 
number of states with ${\cal J}=n/2-1$ and ${\cal J}=n/2$. That is, 
\begin{eqnarray}
g(n,{\cal J}=n/2-2) & = & {n\choose 2}-g(n,{\cal J}=n/2-1)-g(n,{\cal J}=n/2)\nonumber\\
& = & {n\choose 2}-\left[{n\choose 1}-{n\choose 0}\right]
-{n\choose 0}={n\choose 2}-{n\choose 1}, 
\end{eqnarray}    
Continuing
in the same way, we find that
\begin{equation}
g(n,{\cal J})={n\choose n/2-{\cal J}}-{n\choose n/2-({\cal J}+1)}
=\frac{2{\cal J}+1}{m+{\cal J}+1}{2m\choose m+{\cal J}}\;,
\label{deghyp}
\end{equation}  
where $m\equiv n/2$. Note that by setting the hyperspin to zero
Eq.~(\ref{dege}) is obtained. In the Appendix we generalize
Eq.~(\ref{deghyp}) assuming the algebra is constructed from three
building blocks.

The requirement that ${\cal J}=0$ implies that the
states describing the black hole have a definite ``parity''. In our
algebra, the operator $\hat{p}$ that exchanges the two
building blocks is called the parity operator:  
\begin{equation}
\hat{p}|x_{1}x_{2}...x_{n}\rangle\!\rangle
=|\bar{x}_{1}\bar{x}_{2}...\bar{x}_{n}\rangle\!\rangle,  
\end{equation}
where $\bar{x}_{i}=1-x_{i}$ ($i=1,2,...,n$). It can be shown the the
parity operator commutes with $\vec{{\cal J}}^{2}$ but anticommutes with
$\hat{\cal J}_{3}$. Therefore, the only states with definite 
${\cal J}$ and ${\cal J}_{3}$ that have a definite parity are
states with zero hyperspin. Also in the framework of the canonical
quantization of a Schwarzschild black hole~\cite{Vaz} it was shown
that the relevant quantum states that exhibit the equally spaced area
spectrum have a definite parity. Hence, imposing on the quantum states the
requirement of a definite parity is translated in our approach as the requirement
that the hyperspin is zero.     

\section{Discussion}
\label{sec:discussion}
 
Most (and perhaps all) of the observables in physics, such
as momentum, angular momentum and energy arise from
some symmetry transformations. The hyperspin defined in the
previous sections is also derived in the same manner. 
In the algebra addressed in the previous paper~\cite{BekGour}, we have
used two building blocks $\hat{a}$ and $\hat{b}$. The symmetry between 
$\hat{a}$ and $\hat{b}$ implies the existence of the hyperspin.
In harmony with the heuristic approach it is an observable which
represents a global quantum number of the black hole. 

In our algebra we have obtained the ``maximum'' symmetry between 
the two building blocks; i.e. $U(2)$ symmetry. We cannot obtain a 
larger group because there are only two building blocks. In the
Appendix, we discuss the algebra of three building blocks, which,
therefore, has a larger symmetry. It is very natural that in the
construction of an algebra describing the inner structure of a black
hole some other global quantum numbers will appear. The question is
then, how can it live in peace with the ``no hair'' principles?
 
The only way to resolve this issue is to set the hyperspin to
zero. Since we consider here a neutral non-rotating black hole, the
hyperspin cannot represent the angular momentum (or charge) of the
black hole. Further, the entropy will not be a function just of the horizon
area if one considers the hyperspin as the angular momentum. Similar
arguments imply that the
hyperspin cannot represent quantum numbers like charm or
strangeness of ordinary matter; otherwise it
must be very small to be compatible with the classical
limit. Therefore, it is reasonable to assume that the hyperspin
represents a quantum number which does not exist in ordinary
matter. The conservation of the hyperspin then implies that a black
hole which was formed by a gravitational collapse of ordinary matter
must have zero hyperspin. 

Hawking showed very convincingly that at the semi-classical level,
where the gravitational field is treated classically but the matter
fields exterior to the black hole are quantized, black holes radiate
all kinds of particles with spectrum of a black body. If these
particles do not possess hyperspin, but the black hole does, the
Hawking radiation must cease at the point when the horizon area
becomes equal to the hyperspin (see
Eq.~(\ref{areas})). Thus, eternal black holes which do not radiate can have 
${\cal J}\gg 1$. Of course, this hyperspin can not be measured by an
observer outside the black hole.

We conclude that the hyperspin cannot represents a quantum number of
ordinary matter. Instead, it might represent a global feature of 
the $D_{0}$-branes in $M$-theory, the Wilson lines in the Ashtekar
approach, or a quantum number attached to the patches described
in the heuristic approach.

\section{Acknowledgments} 
I would like to thank J.~Bekenstein for his guidance and support.
This research was partly supported by grant No. 129/00-1 of the Israel Science
Foundation and by a Clore Foundation fellowship. The author is also 
grateful for the Killam Trust for its financial support.

\appendix

\section{The algebra of three building blocks}

The algebra of two building blocks~(\ref{algebra1})-(\ref{algebra}) 
can be generalized
to that with three building blocks. Adding another building block, say
$\hat{c}$, one must assume the following commutation relations:
\begin{eqnarray}
(1)\;\;\;[\hat{A},\hat{c}] & = & a_{0}\hat{c}\nonumber\\
(2)\;\;\;[\hat{c}^{\dag},\hat{c}] & = & 1+w\hat{N}\nonumber\\
(3)\;\;\;[\hat{c}^{\dag},\hat{a}] & = & [\hat{c}^{\dag},\hat{b}]=0
\label{calgebra}
\end{eqnarray}
These commutation relations follows from the same arguments that led
to Eq.~(\ref{algebra1}-\ref{algebra}) (see~\cite{BekGour}), and together with 
Eq.~(\ref{algebra1}-\ref{algebra}) constitute the totality algebra of
three building blocks. The algebra is invariant under the transformation:
\begin{equation}
\pmatrix{\hat{a} \cr
\hat{b} \cr \hat{c} \cr}
\;\;\rightarrow\;\;
\pmatrix{v_{11} & v_{12} & v_{13} \cr
v_{21} & v_{22} & v_{23} \cr
v_{31} & v_{32} & v_{33} \cr}
\pmatrix{\hat{a} \cr
\hat{b} \cr \hat{c} \cr}\;\equiv\;
{\bf v}\pmatrix{\hat{a} \cr
\hat{b} \cr \hat{c} \cr}\;,
\end{equation}
where ${\bf v}\in U(3)\equiv U(1)\times SU(3)$ is a 3-dimensional
unitary matrix. The generator of the $U(1)$ symmetry is the area
operator $\hat{N}$ itself. 

The group $SU(3)$ has 8 generators which
are denoted here by $\hat{\cal G}_{k}$ $(k=1,2,...,8)$ (a
basis of the $SU(3)$-Lie algebra generalizing the Pauli matrices is
given by the 8 Gell-Mann matrices~\cite{IZuber}). The commutation 
relations between $\hat{\cal G}_{k}$ $(k=1,2,...,8)$ and $\hat{a}$ (or
$\hat{b}$), can be deduced from
\begin{equation}
\pmatrix{
\exp(-i\theta \hat{\cal G}_{k})\hat{a}\exp(i\theta \hat{\cal G}_{k}) \cr
\exp(-i\theta \hat{\cal G}_{k})\hat{b}\exp(i\theta \hat{\cal G}_{k}) \cr
\exp(-i\theta \hat{\cal G}_{k})\hat{c}\exp(i\theta \hat{\cal G}_{k}) \cr}
=\exp(i\theta {\bf \lambda}_{k})
\pmatrix{\hat{a} \cr \hat{b} \cr \hat{c} \cr}\;,
\label{qnumber}
\end{equation}
where $2\{\lambda_{k}\}$ are the eight Gell-Mann matrices that
generalizing the Pauli matrices~\cite{IZuber}. The first three
generators ${\cal G}_{1}$, ${\cal G}_{2}$ and ${\cal G}_{3}$  satisfy
the same commutation relations given in Eq.~(\ref{commu}) if one
replace ${\cal J}_{k}$ by ${\cal G}_{k}$ for $k=1,2,3$. Therefore, we
shall identify the first three generators with the hyperspin, i.e.
${\cal G}_{k}\equiv {\cal J}_{k}$ for $k=1,2,3$. In addition, from
Eq.~(\ref{qnumber}) it follows that the third building block $\hat{c}$
commutes with the hyperspin ($[\hat{c},\vec{\cal
  J}]=0$). 

Further, Eq.~(\ref{qnumber}) implies that the eight
generator ${\cal G}_{8}\equiv\hat{Y}$ (similar to the hypercharge in
the standard model) commutes with the hyperspin and satisfies the
commutation relations:
\begin{equation}
[\hat{Y},\hat{a}]=\frac{1}{2\sqrt{3}}\hat{a}\;,\;\;
[\hat{Y},\hat{b}]=\frac{1}{2\sqrt{3}}\hat{b}\;\;\;{\rm and}\;\;\;        
[\hat{Y},\hat{c}]=-\frac{1}{\sqrt{3}}\hat{c}.
\end{equation}
That is,
$\hat{Y}=\frac{1}{2\sqrt{3}}(\hat{N}_{a}+\hat{N}_{b}-2\hat{N}_{c})$. 

We can define $3^{n}$ states with the same area just as in
Eq.~(\ref{states}), except that now $x_{i}$ can have the values
$-1,\;0$ or $1$ and correspondingly $\hat{x}_{i}$ is either $\hat{a}$,
$\hat{b}$ or $\hat{c}$. It can be shown that all the $3^{n}$ states
are independent (for two building blocks see the proof
in~\cite{BekGour}). Now, since $\hat{N}$, $\vec{\cal J}^{2}$,
$\hat{\cal J}_{3}$ and $\hat{Y}$ all commute with each other, we
can write the $3^{n}$ black hole states in the form 
$|n,{\cal J},{\cal J}_{3},Y,s\rangle$, where ${\cal J}=0,1,2,...,n/2$ (if
$n$ is even) or ${\cal J}=1/2,3/2,...,n/2$ (if $n$ is odd), 
${\cal J}_{3}=-{\cal J},-{\cal J}+1,...,{\cal J}$,
$\;2\sqrt{3}Y=-n,-n+3,-n+6,...,2n$ and $s=0,1,...,g(n,{\cal J},Y)-1$ 
where $g(n,{\cal J},Y)$ is the number of
states with the same area, hyperspin ${\cal J},\;{\cal J}_{3}$ and
``hypercharge'' $Y$.  

By using the same technique that led to Eq.~(\ref{deghyp}), one can show that
\begin{equation}
g(n,{\cal J},Y)=\frac{2{\cal J}+1}{\eta+y+{\cal J}+1}{3\eta\choose
  \eta+y-{\cal J}}{2\eta-y+{\cal J}\choose \eta-2y}\;,
\label{adeg}
\end{equation}
where $\eta\equiv n/3$ and $y\equiv Y/\sqrt{3}$. Now, since the
hyperspin ${\cal J}$ and also $Y$ are zero (see
Sec.~\ref{sec:discussion}), the degeneracy of the $n$th area level is
given by:
\begin{equation}
g(n)=\frac{1}{1+\eta}{3\eta\choose\eta}{2\eta\choose\eta}\;,   
\end{equation}
where we set ${\cal J}=Y=0$ in Eq.~(\ref{adeg}). Using Stirling's
formula, in the limit $n\rightarrow\infty$
\begin{equation}
g(n)\approx\frac{3^{n}}{n^{2}}.
\end{equation}
Hence, in the case of three building blocks the leading correction
term to the Bekenstein-Hawking entropy is $-2\ln n$.

\end{document}